\documentclass[aps,prd,unsortedaddress,superscriptaddress,showpacs,nofootinbib,twocolumn,10pt]{revtex4-2} 
\usepackage[utf8]{inputenc}
\usepackage{array}
\usepackage{multirow}
\usepackage{graphicx,color}
\usepackage{relsize}
\usepackage{slashed}
\usepackage[normalem]{ulem}
\usepackage{tabu}
\usepackage{rotating}
\usepackage{bigstrut}
\usepackage{makecell}
\usepackage[dvipsnames]{xcolor}
\usepackage[colorlinks=true, linkcolor=blue, citecolor=Green, urlcolor=blue]{hyperref}
\usepackage{amsmath}
\usepackage{amssymb,bm}
\usepackage{amsthm}
\usepackage{mathrsfs}
\usepackage{array}
\usepackage[all]{xy}
\usepackage{euscript}
\usepackage{enumerate}
\usepackage{mathtools}
\usepackage{soul}
\usepackage{orcidlink}

\newcommand{\dsz}{D_{s0}^*(2317)}
\newcommand{\dsone}{D_{s1}(2460)}

\allowdisplaybreaks

\graphicspath{{Figs/}}  

\newcommand{\itp}{\affiliation{CAS Key Laboratory of 
Theoretical Physics, Institute of Theoretical Physics,\\
Chinese Academy of Sciences, Beijing 100190, China}}
\newcommand{\ucas}{\affiliation{School of Physical Sciences, 
University of Chinese Academy of Sciences, Beijing 100049, 
China}}
\newcommand{\peng}{\affiliation{Peng Huanwu Collaborative 
Center for Research and Education, Beihang University, Beijing 
100191, China}}

\newcommand{\scnt}{\affiliation{Southern Center for Nuclear-Science Theory (SCNT), Institute of Modern Physics,\\ 
Chinese Academy of Sciences, Huizhou 516000, China}}

\newcommand{\fzj}{\affiliation{Institute for Advanced Simulation (IAS-4), Forschungszentrum J\"ulich, D-52425 J\"ulich, Germany}}

\begin{document}

\title{
Are compact open-charm tetraquarks consistent with recent lattice results? 
}

\author{Feng-Kun~Guo\orcidlink{0000-0002-2919-2064}}
\email{fkguo@itp.ac.cn}
\itp\ucas\peng\scnt

\author{Christoph Hanhart\orcidlink{0000-0002-3509-2473}}\email{c.hanhart@fz-juelich.de}
\fzj

\begin{abstract}
We argue that the hypothesis that positive-parity charm meson resonances exhibit a compact tetraquark structure has some clear tension with recent lattice results for the $S$-wave $\pi D$ system for an SU(3) flavor symmetric setting.
In particular, we show that 
such a diquark--anti-diquark tetraquark scenario would call for the presence of a state
in the flavor $[{\mathbf{\overline{15}}}]$ representation, not seen in the lattice analysis.
Moreover, we show that analogous lattice data in the
axial-vector channel are even more sensitive to the
internal structure of these very interesting states.
\end{abstract}

\maketitle

\section{Introduction}
\label{Sec:intro}

Since their discovery in 2003 the lightest open-charm positive-parity mesons containing strangeness remained largely a mystery---especially in light of the seemingly expected
properties of their non-strange partner states. In the past, attempts were made to explain the low-lying charm-strange $D_{sJ}$ states as $c\bar s$ mesons~\cite{Cahn:2003cw,Godfrey:2003kg,Colangelo:2003vg,Mehen:2005hc,Lakhina:2006fy}, chiral partners of the ground state $D_s$ and $D_s^*$ mesons~\cite{Bardeen:2003kt,Nowak:2003ra}, compact $[cq][\bar s\bar q]$ tetraquark states~\cite{Maiani:2004vq}, mixing of the $c\bar s$ and tetraquarks~\cite{Browder:2003fk}, a $D\pi$ atom for the $\dsz$~\cite{Szczepaniak:2003vy}, and $D^{(*)}K$
hadronic molecules~\cite{Barnes:2003dj,vanBeveren:2003kd,Chen:2004dy,Kolomeitsev:2003ac,Guo:2006fu,Guo:2006rp, Gamermann:2006nm}.
The experimental data show four features that need to be understood~\cite{Du:2017zvv}:
\begin{enumerate}\setlength\itemsep{-0.3em}
  \item[(1)] {\it The $D_{sJ}$ states are too light}: Both $D_{s0}^*(2317)$ and $D_{s1}(2460)$ are much lighter than the quark model expectations~\cite{Godfrey:1985xj} for the lowest scalar and axial-vector $c\bar s$ mesons.
  \item[(2)] {\it Fine-tuning}: Within only 2~MeV the relation $M_{\dsone}{-}M_{\dsz}{\approx} M_{D^{*\pm}}-M_{D^\pm}$ holds. 
  \item[(3)] {\it Mass hierarchy}: One finds $M_{D_0^*(2300)}\sim M_{D_{s0}^*(2317)}$
   and $M_{D_1(2430)}\sim M_{D_{s1}(2460)}$, although usually adding a strange quark leads to an increase in mass of about 150-200 MeV.
    \item[(4)] {\it Discrepancy theory versus experiment}: Various theoretical studies~\cite{Kolomeitsev:2003ac,Guo:2006fu,Guo:2006rp,vanBeveren:2003kd,Albaladejo:2016lbb}, later confirmed by lattice QCD~\cite{Gayer:2021xzv,Lang:2022elg}, find lower masses for $D_0^*$ and $D_1$ than values reported from experiments~\cite{ParticleDataGroup:2024cfk}.
\end{enumerate}

All these find a natural explanation, if the lowest positive-parity charmed mesons are interpreted as hadronic mole\-cules~\cite{Du:2017zvv}.
In this case, the flavor structure of this family of states is governed by  the flavor decomposition representing the scattering of the lightest pseudoscalar mesons off $D$ mesons, which are SU(3) flavor octets and anti-triplets, respectively. This decomposition is given by~\cite{Kolomeitsev:2003ac, Albaladejo:2016lbb}
\begin{equation}
    [\mathbf{\bar 3}]\otimes [\mathbf{8}] = [\mathbf{\bar 3}] \oplus[\mathbf{6}]\oplus
[{\mathbf{\overline{15}}}] \ .
\label{eq:mol_decomp}
\end{equation}
The light pseudoscalar octet corresponds to the pseudo-Nambu-Goldstone (pNG) bosons of the spontaneously broken chiral symmetry of quantum chromodynamics (QCD). 
Note that we do not include scattering of the ninth pseudoscalar, the $\eta'$, here, since due to the action of the U$(1)_A$ anomaly it is not a pNG boson. 

Non-strange isospin-1/2 multiplets appear in all three irreducible representations, however, chiral symmetry constraints dictate
that only the $[\mathbf{\bar 3}]$ and the $[\mathbf{6}]$ are attractive for the charm-light meson scatterings~\cite{Kolomeitsev:2003ac, Albaladejo:2016lbb}. 
In particular,
in this scheme the $D_0^*(2300)$ listed in the Review of Particle Physics~\cite{ParticleDataGroup:2024cfk}
is interpreted as emerging from
 two distinct poles (resonances), one at 2105~MeV and the other at 2451~MeV, with the lower one being a member of the same  SU(3) multiplet as the $D_{s0}^*(2317)$, the $[\mathbf{\bar 3}]$, where the attraction is the strongest.  The state at 2451~MeV is a member of the $[\mathbf{6}]$ representation of SU(3), where the interaction is weaker than in the $[\mathbf{\bar 3}]$, but still sufficiently strong to generate a pole at physical quark masses sufficiently close to the physical axis to show an impact on observables~\cite{Du:2017zvv,Du:2020pui}. 
Chiral symmetry constraints impose that the leading order interaction of the particles in the $[{\mathbf{\overline{15}}}]$ is repulsive and thus no resonance should be found in this channel. 

Thus, a crucial test of this interpretation is connected to the existence of the $\mathbf{[6]}$ state with a pole located around 2451~MeV
and the absence of a pole in
the $[{\mathbf{\overline{15}}}]$.
In contrast to this, conventional quark model states with a quark composition $c\bar u$ or $c\bar d$ can appear only in the flavor $[\mathbf{\bar 3}]$
representation. 

One way to unravel the SU(3) structure 
underlying the spectrum of the lightest open charm scalar states
is to unambiguously establish the existence of the state in the 
$\mathbf{[6]}$ by determining in the SU(3) symmetric limit whether it appears as a near-threshold pole, when the pNG boson mass (i.e. the pion mass) is sizably larger than the physical value, as predicted by unitarized chiral perturbation theory~\cite{Liu:2012zya,Du:2017zvv}. This needs to be accompanied by the absence of
a state in the $[{\mathbf{\overline{15}}}]$.
First results indicated that, indeed, the low-energy charm-light meson scattering in the $\mathbf{[\overline{15}]}$ is \emph{repulsive} and that in the $\mathbf{[6]}$ is \emph{attractive} with a near-threshold pole, thus providing strong evidence for the molecular nature for these states~\cite{Gregory:2021rgy}.
This finding was confirmed recently~\cite{Yeo:2024chk}
by a detailed  L\"uscher analysis imposing similar quark masses.

While these findings provide strong support for a molecular structure
of the mentioned states, it remains to be studied what the
compact tetraquark picture predicts for this system, with
building blocks being colored (anti)-diquarks (for a review of the concept see 
Ref.~\cite{Barabanov:2020jvn}). 
In particular, in this work we critically review the claim of Ref.~\cite{Maiani:2024quj} that there exists a scenario
where compact tetraquarks also appear only in
the flavor $[\mathbf{\bar 3}]$ and $[\mathbf{6}]$ representations
and not in the $[{\mathbf{\overline{15}}}]$.
In contrast to this claim, we argue that as soon as one allows for axial-vector diquarks in addition to the scalar diquarks investigated in Ref.~\cite{Maiani:2024quj} additional states in the $[{\mathbf{\overline{15}}}]$ emerge.
This observation appears to be already at odds with the findings in lattice QCD quoted above. Moreover, we demonstrate that additional data at similarly high unphysical quark masses
in the axial-vector channel should clearly show a signature of a state in the $[{\mathbf{\overline{15}}}]$ representation if the compact tetraquark picture were correct, while no such state should appear in this multiplet in the molecular picture. We are thus close to unambiguously identifying the structure of heavy-light multi-quark states.

\section{Flavor structure of light and heavy light diquarks}

A detailed discussion
of the SU(3) flavor structure of  positive-parity open-heavy-flavor mesons within the
tetraquark picture can be found in Refs.~\cite{Dmitrasinovic:2004cu,Dmitrasinovic:2005gc,Maiani:2024quj,Dmitrasinovic:2023eei}. 
To be concrete, consider a tetraquark state with a heavy quark, a light quark and two light antiquarks. The flavor decomposition of the light degrees of freedom reads
\begin{eqnarray}\nonumber
[\mathbf{3}]\otimes[\mathbf{\bar 3}]\otimes[\mathbf{\bar 3}]&=& [\mathbf{3}] \otimes( [\mathbf{3}_A] \oplus [\mathbf{\bar 6}_S]) \\ 
&=& [\mathbf{\bar 3}_A] \oplus [\mathbf{\bar 3}_S] \oplus [\mathbf{6}_A]\oplus
[{\mathbf{\overline{15}}}_S] \ ,
\label{eq:tetraquarkdecomp}
\end{eqnarray}
where the labels $A$ and $S$ indicate that in these flavor multiplets the two antiquarks are antisymmetric or symmetric in flavor space, respectively. 
Note that in contrast to Eq.~(\ref{eq:mol_decomp}), which contains 24 states in total, here 27 states appear. This difference arises because the $\eta'$ was 
not included in the flavor decomposition of Eq.~(\ref{eq:mol_decomp}), as mentioned above.
We will return to this feature below.

In Refs.~\cite{Dmitrasinovic:2005gc, Dmitrasinovic:2004cu}, it is argued that the observed states are tetraquarks in the flavor $[{\mathbf{\bar 3}}]$ and $[{\mathbf{\overline{15}}}]$ representations---features that were deduced by employing the 't~Hooft interaction at the diquark level. This scenario is clearly ruled out by the lattice data quoted above. However, these works do not discuss the role of the diquark spin.

In contrast to the studies in Refs.~\cite{Dmitrasinovic:2005gc, Dmitrasinovic:2004cu}, Ref.~\cite{Maiani:2024quj} takes as its key starting point that the Pauli principle imposes restrictions on the light (anti)diquark quantum numbers (see, e.g., Ref.~\cite{Jaffe:2004ph}). Specifically, it emphasizes that an $S$-wave diquark in the color anti-triplet with zero angular momentum must have the same symmetry in both spin and flavor. When focusing on the spin-zero diquark, this implies that only the two flavor antisymmetric multiplets for tetraquarks, $[\mathbf{\bar 3}_A]$ and $[\mathbf{6}_A]$, are allowed, consistent with the findings of the lattice studies in Refs.~\cite{Gregory:2021rgy,Yeo:2024chk}.
To accommodate both $D_{s0}^*(2317)$ and $D_{s1}(2460)$ in this picture, the mass difference between the spin-one and spin-zero $cq$ diquarks, which are partners under heavy quark spin symmetry (HQSS), must approximately equal that of $D^*$ and $D$. This requirement is not only phenomenologically sensible but also addresses the fine-tuning issue mentioned in the Introduction.
Therefore, one has, at the physical point, 
\begin{eqnarray}\nonumber
    M_{cq}^{[S=1]}- M_{cq}^{[S=0]}&\approx&M_{D_{s1}(2460)}-M_{D_{s0}^*(2317)} \\ &=& 142 \ \mbox{MeV} \ .
    \label{eq:cqmassdifference}
\end{eqnarray}
This spin symmetry violating mass difference should be only
weakly dependent on the pion mass due to the $1/m_c$ suppression, where $m_c$ is the charm quark mass, and thus we will use it also
in estimates for mass differences at heavier pion masses below.

The construction scheme of diquark models implies
that
not only the flavor $[\mathbf{3}_A]$ diquark in $S$-wave lives in the color anti-triplet, but also the $[\mathbf{\bar 6}_S]$ diquark, and thus both possess an attractive one-gluon exchange potential.
If the spin-one $cq$ diquark exists as a relevant degree of freedom, as required by a consistent diquark phenomenology, the flavor blindness of gluonic interactions would imply the existence of a spin-one light $qq$ diquark (and $\bar q \bar q$ anti-diquark) as well, since the exchange-momentum scales for both systems are similar, $\sim\Lambda_{\rm QCD}$.
Since here the spin wave function is symmetric, it must be combined with the flavor symmetric multiplets for anti-diquarks in the first line of Eq.~(\ref{eq:tetraquarkdecomp}). Consequently, spin-one $\bar q\bar q$ anti-diquarks appear in the $[\mathbf{\bar 3}_S]$
and $[{\mathbf{\overline{15}}}_S]$ representations for tetraquarks in the second line of Eq.~\eqref{eq:tetraquarkdecomp}. The mass of the anti-diquark contained in these can be
estimated from the $\Sigma_c$-$\Lambda_c$ mass difference. Thus one obtains
\begin{eqnarray}
    M_{\bar q\bar q}^{[S=1]}- M_{\bar q\bar q}^{[S=0]}&\approx&M_{\Sigma_c}-M_{\Lambda_c} \notag\\ &=& 167 \ \mbox{MeV} \ 
    \label{eq:spin1mass}
\end{eqnarray}
for physical quark masses. 
HQSS predicts that this mass difference needs to agree with that of the respective bottomed baryons (with an uncertainty of about 20 MeV due to relative corrections of order $\Lambda_{\rm QCD}(1/m_c-1/m_b)$), which is 191~MeV lying in the expected range.
It also agrees with the estimate $\sim \frac{2}{3}\left(M_{\Delta}-M_N\right) \approx 180~\mathrm{MeV}$ in Ref.~\cite{Jaffe:2004ph}.

Given both spin-zero and spin-one diquarks and these mass splittings, it should
be expected that additional multiplets develop bound states in the tetraquark picture, besides the ones considered in Ref.~\cite{Maiani:2024quj}.
In the next section we
argue that states of the $[\mathbf{\overline{15}}_S]$ representation should be present
in the scalar and especially the axial-vector 
sector of the open charm states, if
those states had a tetraquark structure of the diquark--anti-diquark type.

\section{Results}

In Ref.~\cite{Maiani:2024quj}, it is argued that the states in the flavor sextet and those in the anti-symmetric flavor anti-triplet should be near mass degenerate. However, this is in conflict with the results of Ref.~\cite{Yeo:2024chk}, at least at the unphysically large quark mass used in the lattice calculation. 
On the one hand, a state in the flavor anti-triplet was found in the lattice study as a deeply bound state at a mass of about 2435~MeV, corresponding to a binding energy of slightly above 200~MeV, at a pion mass of about 700~MeV in a flavor symmetric setting. 
On the other hand, the flavor sextet state appeared as a virtual state with a pole between 2510 and 2610 MeV.\footnote{The lattice results qualitatively agree with the prediction in the molecular picture~\cite{Du:2017zvv}.} Although this pole lies 40-140 MeV below threshold, it should not be directly compared with the mass of a bound state having a similar binding energy.
This is because a bound state indicates strong attraction in the respective channel---when the attraction is reduced, the bound state first moves to the threshold and only after that it becomes a virtual state.

In this work we want to investigate to what extent the lattice data are consistent with the compact tetraquark picture. To that end we need to first of all assume that the state in the sextet is a very shallow bound state and not a virtual state, since the latter property would point at a molecular structure~\cite{Matuschek:2020gqe}, although that introduces already some tension to the lattice data.

Since the mentioned lattice analysis was performed in a flavor symmetric setting, there is no mixing between states in different flavor multiplets and the mass difference between them thus directly reflects the mass difference between the multiplets.
We thus have (at least at $M_\pi\approx700$~MeV)
\begin{equation}
    M_{[\mathbf{6}_A]} -  M_{[\mathbf{\bar 3}_A]} \approx 200 \ \mbox{MeV}  .
    \label{eq:36split}
\end{equation}
This observation puts into 
question the quantitative analysis of Ref.~\cite{Maiani:2024quj}.

In Ref.~\cite{Dmitrasinovic:2004cu}, a mechanism is presented that naturally provides a mass splitting between the mentioned multiplets driven by the 't Hooft interaction. In particular, it is argued that
the 't Hooft interaction induces mass shifts that lift the naively expected
degeneracy between the multiplets (see Eqs.~(12-15) therein), in particular the $[\mathbf{\bar 3}_A])$ 
acquires a rather strong additional attraction,
both $[\mathbf{\overline{15}}_S]$ and $[\mathbf{6}_A]$
a milder one, while the $[\mathbf{\bar 3}_S]$
feels a rather strong additional repulsion. 

These considerations let us ignore the states
in the $[\mathbf{\overline{3}}_S]$ in what follows.
Then we find, using the estimate of  
Eq.~(\ref{eq:spin1mass}) for the mass difference of the spin 1 and the spin 0 diquark, the mass of the lightest axial-vector tetraquark state in the $[\mathbf{\overline{15}}_S]$ representation with spin-zero $cq$ diquark and spin-one light anti-diquark to be
\begin{equation}
    M_{[\mathbf{\overline{15}}_S],1^+} \approx 2600 \ \mbox{MeV} ,
    \label{eq:mass15spin1}
\end{equation}
which refers to a deep bound state with a binding
energy of about 50 MeV, at the SU(3) symmetric point with a 700~MeV pion mass.

The axial-vector mass in Eq.~(\ref{eq:mass15spin1}) is very close to the mass of the axial-vector state in the flavor $[\mathbf{\overline{3}}_A]$ representation, which we
find from adding the mass difference between the spin-one and spin-zero $cq$ diquark, 
Eq.~(\ref{eq:cqmassdifference}), to the mass of the lightest scalar found in the lattice study to find
\begin{equation}
    M_{[\mathbf{\overline{3}}_A],1^+} \approx 2575 \ \mbox{MeV} .
    \label{eq:massbar3spin1}
\end{equation}
For this estimate we assumed
that the mass difference between the spin-one and spin-zero $cq$ diquarks at the SU(3) symmetric point with a large pion mass is similar to the one at the physical point.
We therefore conclude
that in the compact tetaraquark picture their should be a pair of close
by deep bound states in the axial vector sector, one in the flavor
$[\mathbf{\bar 3}_A]$ and one in the flavor $[\mathbf{\overline{15}}_S]$.
At the same time, the 
states in the $[\mathbf{\bar 3}_S]$ should be pushed
so high up that they do not matter in the region of interest---this is in fact analogous to the fate of the $\eta'$ in the mesonic
analysis.

As argued above, the anti-diquark in the
$[\mathbf{\overline{15}}_S]$ should have spin one. Thus, combining this with a spin-zero $cq$ diquark one gets axial vectors for the states given in Eq.~(\ref{eq:mass15spin1}) above. On the other hand,
combining it with a spin-one $cq$ diquark, one gets three different possible total spins: 0, 1 and 2. All these states are grouped into two different multiplets under HQSS: $\frac{1}{2}_q \otimes 1_{\{\bar{q}\bar{q}\}}= \left(\frac{1}{2} \oplus \frac{3}{2}\right)_{q\{\bar{q} \bar q\}}$ for the total angular momentum of the light degrees of freedom $j_\ell$, where we use ${\{\bar{q}\bar{q}\}}$ to denote the flavor symmetric anti-diquark. Coupling this to the heavy quark spin $s_c=1/2$, one gets two tetraquark multiplets under HQSS: one is given by 
\begin{align}
    \frac{1}{2}_c \otimes \frac{1}{2}_{q\{\bar{q} \bar q\}}=\left(0 \oplus 1\right)_{(cq)\{\bar{q} \bar q\}}, \label{eq:spin1D-01}
\end{align}
and the other is given by 
\begin{align}
    \frac{1}{2}_c \otimes \frac{3}{2}_{q\{\bar{q} \bar q\}}= \left(1 \oplus 2\right)_{(cq)\{\bar{q} \bar q\}}. \label{eq:spin1D-12}
\end{align}
Thus, there should be several $[\mathbf{\overline{15}}_S]$ in SU(3) flavor space, and they have different total spins: $$(0\oplus 1)_{cq\{\bar{q} \bar q\}} \oplus (1 \oplus 2)_{cq\{\bar{q} \bar q\}}\ ,$$ where each HQSS spin doublet is paired inside parentheses. 

Furthermore,
the HQSS spin multiplets of charmed tetraquarks containing a spin-zero light anti-diquark, being antisymmetric in flavor and denoted by $[\bar{q} \bar q]$, read 
\begin{align}
    \frac{1}{2}_c \otimes \frac{1}{2}_q\otimes 0_{[\bar{q} \bar q]}=\left(0 \oplus 1\right)_{(cq)[\bar{q} \bar q]} \ . \label{eq:spin0D}
\end{align}
This spin decomposition holds for both flavor $[\mathbf{\overline{3}}_A]$ and $[\mathbf{6}_A]$.

To find the mass
of the corresponding scalar $[\mathbf{\overline{15}}_S]$ state in Eq.~\eqref{eq:spin1D-01}, we need to 
add the mass in Eq.~(\ref{eq:mass15spin1}) to the mass difference between the spin-zero and
spin-one $cq$ diquarks provided in Eq.~(\ref{eq:cqmassdifference}). 
Putting all pieces together we estimate (neglecting
the hyperfine splitting for the couplings of spin-one $cq$ and spin-one $\bar q\bar q$ in line
with other studies in this context, e.g., Ref.~\cite{Maiani:2004vq}) 
\begin{equation}
    M_{[\mathbf{\overline{15}}_S],0^+} \approx 2740 \ \mbox{MeV} ,
    \label{eq:mass15spin0}
\end{equation}
which means that this naive estimate for the scalar $[\mathbf{\overline{15}}_S]$ state would have a mass higher than the SU(3) symmetric threshold, about 2650~MeV.
Thus it should probably show up as a resonance.  However, such a state was not found in the lattice
analysis of Ref.~\cite{Yeo:2024chk}.

\begin{figure}
    \centering
\includegraphics[width=\linewidth]{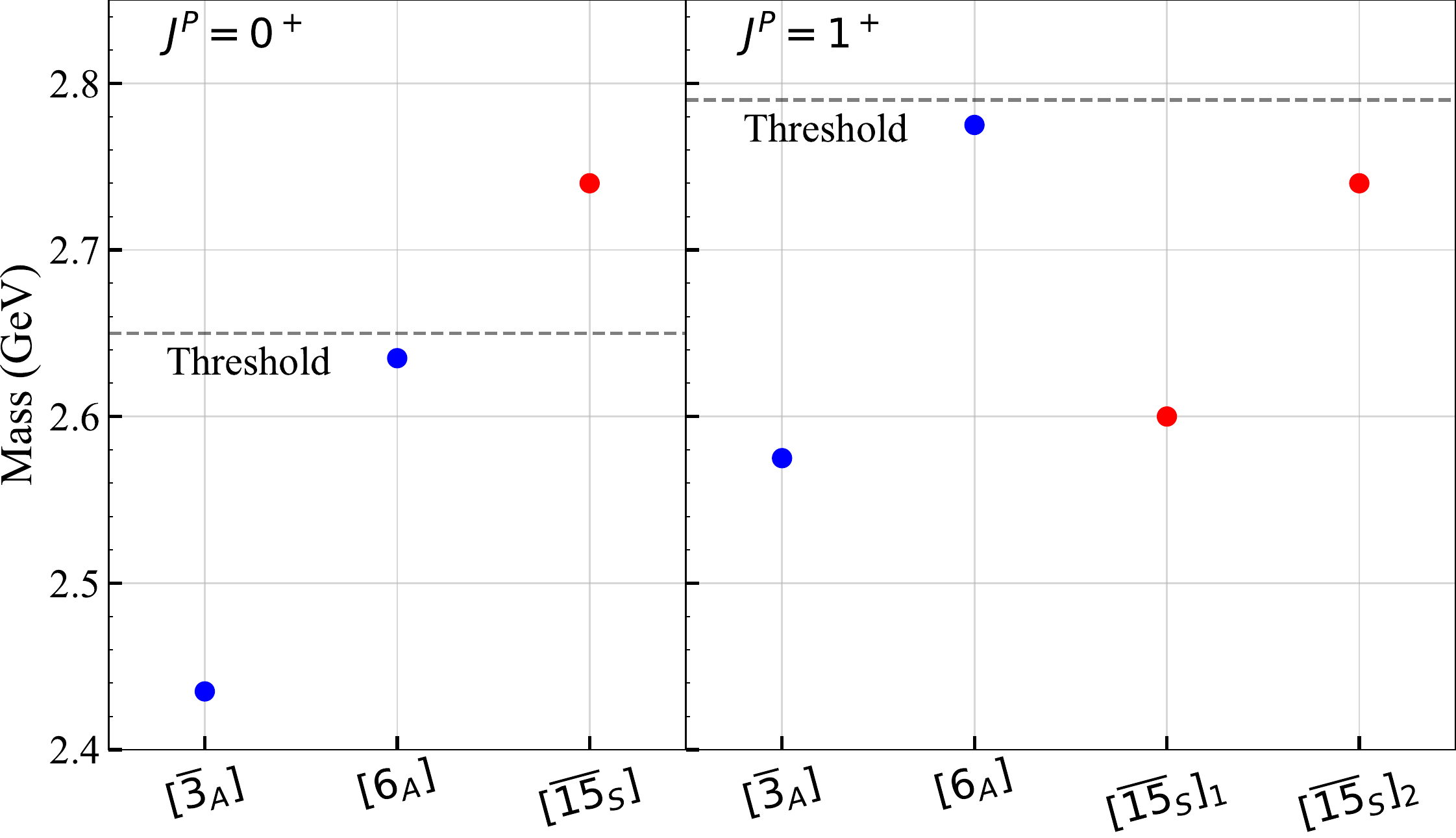}
    \caption{Sketch of the spectrum of scalar and axial-vector tetraquarks in the diquark--anti-diquark model
    for a pion mass of about 700 MeV.
    Blue (red) symbols refer to states containing spin-zero (spin-one) light diquarks.
    In contrast, in the molecular picture, only the analogs
    of the blue states appear with the higher one being located on the unphysical sheet.}
    \label{fig:spec}
\end{figure}
The mass spectrum for the scalar and axial-vector tetraquarks in the diquark--anti-diquark picture is summarized in Fig.~\ref{fig:spec}.
Our analysis shows that in the
tetraquark picture the axial-vector spectrum should look strikingly different from the scalar spectrum of the lightest positive-parity open-charm mesons. 
Although the masses quoted above must be regarded as rough estimates given the intrinsic uncertainties for the individual ingredients and the neglected interactions, qualitatively this observation appears unavoidable, if
(anti)diquarks can be regarded relevant degrees of freedom 
in different types of hadrons.
Clearly, the spectrum shown in Fig.~\ref{fig:spec} is very much different to what is predicted in the molecular picture, where the spectra in the scalar and axial-vector sector are basically the same, because of HQSS for charmed-meson--light-meson interactions, and there is no state present
in the flavor $[\mathbf{\overline{15}}_S]$.

Thus, given the lattice data presently available
in the scalar sector, no strong case against the tetraquark picture can be made considering only the masses. 
Nevertheless, it is important to notice that not only the masses but also the finite-volume energy levels computed in lattice QCD and the $D\pi$ invariant mass distributions in $B$ decays measured by LHCb can be well described in the molecular picture (for a recent concise review, see Ref.~\cite{Guo:2023wkv}).
A description of such detailed data in the diquark--anti-diquark tetraquark picture is still missing. 

Here, it would also be very important to get data in the axial-vector channel. There again the lightest state should be in the $[\mathbf{\bar 3}_A]$
multiplet emerging from the spin-one $cq$ diquark in combination with the spin-zero light anti-diquark. However, the same quantum numbers can be reached by combining the spin-zero $cq$ diquark with the spin-one light anti-diquark, as mentioned above. Taking all the different pieces
together these two states should have roughly the same mass in the compact tetraquark model.
Recently, there have been lattice results in this channel with $D^*\pi, D^*\eta$ and $D_s^*\bar K$ coupled channels at a 391~MeV pion mass~\cite{Lang:2025pjq}. Therein, the lightest state was found just below the $D^*\pi$ threshold, and one additional state arises despite of large uncertainties. This two-pole structure is consistent with the results in Ref.~\cite{Du:2017zvv} (for a general discussion on the two-pole structures, see Ref.~\cite{Meissner:2020khl}). Yet, no state that could be assigned to the 15-plet was found in the lattice analysis.

\section{Summary}

In this paper we argue that especially the recent lattice data for open-charm positive-parity mesons with spin zero at a flavor symmetric point with a pion mass of the order of 700 MeV show some tension with predictions that would emerge from a compact tetraquark picture, while they are in nice agreement with expectations from the molecular ansatz.

It is also demonstrated that the differences between the predictions of the two structure assumptions get 
a lot more drastic when moving to the axial-vector channel. In the diquark--anti-diquark tetraquark picture a pair of two lightest states should emerge: one in the flavor $[\mathbf{\overline{3}}_A]$ with a spin-one $cq$ diquark combined with a spin-zero light anti-diquark, and one in flavor $[\mathbf{\overline{15}}_S]$ with a spin-zero $cq$ diquark combined with a spin-one light anti-diquark. 
Both states should appear as deeply bound states with a mass difference of less than 50~MeV.
Higher up in energy, there should then be a signal in the flavor $[\mathbf{6}_A]$ as well as another one in the $[\mathbf{\overline{15}}_S]$.
Again, relatively close together, although here the uncertainty is even larger.

In the molecular picture, there should be only
one deeply bound state, namely 
the anti-triplet state, also for the axial-vector system showing a binding energy close to what was found in the scalar sector, a virtual state
in the flavor $[\mathbf{6}]$, and no pole at all in the $[\mathbf{\overline{15}}]$.\footnote{At higher energies, there could be more states generated with the pNG bosons replaced by the light vector mesons~\cite{Molina:2010tx,Du:2017zvv,Guo:2020INT}.}

We thus conclude that additional lattice data in the axial-vector channel with the same lattice setting as already employed in Ref.~\cite{Yeo:2024chk} would be extremely valuable in learning whether the lightest positive-parity charmed mesons can be classified as hadronic molecules or compact tetraquarks.

\begin{acknowledgments}
We would like to thank Ulf-G.~Mei{\ss}ner for a careful reading of the manuscript and Luciano Maiani for comments.
This work is supported in part by the National Natural Science Foundation of China (NSFC) under Grants No. 12125507, No. 12361141819, and No. 12047503; by the National Key R\&D Program of China under Grant No. 2023YFA1606703; and by the Chinese Academy of Sciences (CAS) under Grants No.~YSBR-101. In addition, C.H. thanks the CAS President's International Fellowship Initiative (PIFI) under Grant No. 2025PD0087 for partial support. 
\end{acknowledgments}

\bibliography{refs}

\end{document}